\documentstyle[aps,prl,twocolumn]{revtex}
\draft
\begin{document}
\title{
A Nonperturbative Perspective on Inner Product Quantization: Highly Accurate
Solutions to  
the Schr{\"o}dinger Equation}
\author{C. J. Tymczak, G. S. Japaridze, C. R. Handy and Xiao-Qian Wang}
\address{Department of Physics \& Center for Theoretical Studies of 
Physical Systems, Clark Atlanta University, 
Atlanta, Georgia 30314}
\date{Received \today}
\maketitle
\begin{abstract}
We devise a new and highly accurate quantization procedure for the inner 
product representation, both in configuration and momentum space.
Utilizing the representation 
$\Psi(\xi) = \sum_{i}a_i[E]\xi^i R_{\beta}(\xi)$, 
for an appropriate
reference function, $R_{\beta}(\xi)$,
we demonstrate that the (convergent) zeroes of the coefficient 
functions, $a_i[E] = 0$,
approximate the exact bound/resonance state energies
with increasing accuracy as $i \rightarrow \infty$. 
The validity of the approach is shown to be based on an extension
of the Hill determinant quantization procedure. 
Our method has been applied, with remarkable success, 
to various quantum mechanical problems.
\end{abstract}
The development of numerically efficient algorithms for quantum 
mechanical problems
continues to be a challenging issue. Within this context,
variational approaches based on optimal 
selection of basis sets remain
one of the more successful approaches. Nevertheless, it appears that little
attention has been paid to further enhancing such techniques by utilizing
basis functions with an inherently analytical (i.e., power series)
significance. In this Letter we
show that such considerations can greatly reduce the computational effort
of variational procedures by significantly 
reducing the size of the problem.
Many important quantum mechanical problems, heretofore involving large
variational matrices, reduce to solving for the roots of an 
algebraically generated function. 
Our approach is based on a relatively simple extension
of the Hill determinant approach, which nevertheless, 
has not been exploited until now.

One of the basic procedures for solving quantum systems
is the Hill determinant approach [1]. It
involves approximating the bound state wave
function in terms of a suitable truncated basis,
$\Psi(x) =  \sum_{i=0}^I v_i{\cal B}_i(x)$,
and solving the finite dimensional problem
\begin{equation}
\sum_{j=0}^{I}{\cal M}_{i,j}[E] v_j = 0,
\end{equation}
through the Hill determinant equation
\begin{equation}
Det\Big({\cal M}^{(I)}[E]\Big) = 0,
\end{equation}
${\cal M}^{(I)}_{ij} = {\cal M}_{ij}$, for $0 \leq i,j \leq I$, and
${\cal M}_{ij}[E] =  \langle{\cal B}_i|H|{\cal B}_j\rangle -
E \langle{\cal B}_i|{\cal B}_j\rangle$. 
For a suitable basis, as
$I \rightarrow \infty$, the roots of the Hill determinant converge
to the true eigenvalues of the Hamiltonian.

Another basic analytical tool 
is the power series expansion method,
specifically in the form of the inner product representation [2]:
\begin{equation}
\Psi(x) = (\sum_{i}a_i[E]x^i) R_{\beta}(x),
\end{equation}
for an appropriate reference function $R_{\beta}(x)$. 
The function coefficients, $a_i[E]$, depend on the energy variable, $E$, 
and are readily generated through the standard methods for 
differential equations. 
For simplicity, the above expansion assumes 
that $x = 0$ is a regular point. 
If the functions $x^i R_{\beta}(x)$ define a complete basis 
(not necessarily orthonormal),
one can  also pursue a Hill determinant
analysis for the corresponding representation
\begin{equation}
\Psi(x) = \sum_{i}v_ix^iR_{\beta}(x).
\end{equation}

In some cases, the recursive 
structure of the Hill determinant for increasing 
values of $I$ can be computed [3]. 
This allows one to analyze the asymptotic behavior with respect to $I$, 
for the roots of the Hill-determinant equation. 
In general, this analysis can be difficult 
and computationally demanding. 
It is in  this context that we have discovered a remarkable
relation whose simplicity has apparently gone 
unrecognized until now, despite the
suggestive, but specialized, nature of the work by Bender and Dunne [4,5]. 
Specifically, we demonstrate that the convergent zeroes
of the coefficient functions:
\begin{equation}
a_{i+1}[E^{(i)}_l] = 0, 
\end{equation}
(where $l$ lables the roots) 
converge to the exact discrete state energies, $E^{\rm (exact)}_l$, as 
the expansion order, $i$, increases:
\begin{equation}
\lim_{i \rightarrow \infty} E^{(i)}_l = E^{\rm (exact)}_l\ .
\end{equation}

The  theoretical justification for this proceeds as follows. Assume that
${\cal B}_i(x) = x^iR_{\beta}(x)$ and that
the corresponding Hill
determinant method yields convergent results to the 
physical energies and wave functions. Let $E$ assume any 
value, $E = E_{c}$, for which the
infinite matrix ${\cal M}_{ij}[E_c]$,  
has no minor sub-matrix 
with zero determinant. One can
recursively generate [6], through an effective 
LU decomposition method, an infinite set of vectors
$\{V^{(I)}|0 \leq I < \infty\}$ 
%(whose implicit $E_c$ dependence is assumed) 
satisfying, 
\begin{equation}
\sum_{j=0}^I{\cal M}_{i,j}[E_c] \,\, V^{(I)}_j = 0,
\end{equation}
for $0 \leq i \leq I-1$, and
\begin{equation}
\sum_{j=0}^I{\cal M}_{I,j}[E_c] \,\, V^{(I)}_j  = {\cal D}_I[E_c],
\end{equation}
where $  V^{(I)}_I = 1, {\rm and}\ V^{(I)}_j = 0, {\rm for}\ j \geq I+1.$
One also has 
$Det\Big( {\cal M}^{(I)}[E_c] \Big) = \Pi_{i=0}^I {\cal D}_i[E_c]$. 
The relation in Eq. (7) involves $I$ constraints for $I$ unknowns 
(recall $V^{(I)}_I = 1$, thus  Eq. (7) 
is actually an inhomogeneous relation). The second relation,
Eq. (8), serves to define ${\cal D}_I[E_c]$. 

For a given order $I$, the roots of Eq. (2)  
corresponds to the roots of Eq. (8), $E = E_l^{(I)}$, defined by 
$Det\Big( {\cal M}^{(i < I)}[E_{l}^{(I)}]\Big) \neq 0$ and
$Det\Big( {\cal M}^{(I)}[E_{l}^{(I)}]\Big) =  
{\cal D}_I[E_{l}^{(I)}] = 0$. We denote
the corresponding vectors by $V^{(I)}[E_{l}^{(I)}]$.

From the recursion formula for the $V$'s [6], we have:
\begin{equation} 
V^{(I+1)}_{I}[E_{l}^{(I)}] = 
-{{\sum_{i=0}^{I} V^{(I)}_{i}[E_{l}^{(I)}] {\cal M}_{i,I+1}[E_{l}^{(I)}]}
\over
{\cal D}_I[E_{l}^{(I)}]}\ .
\end{equation}
Thus, in the $E \rightarrow E_{l}^{(I)}$ limit, for a given $l$, one obtains 
\begin{equation}
V^{(I+1)}_{I}[E_{l}^{(I)}] = \pm \infty \ .
\end{equation}

By the very nature of the Hill determinant approach, 
for a sufficiently large order, $I+1$, one
expects the partial sums
\begin{equation}
\Psi_l(x) \approx \sum_{i=0}^{I+1} V^{(I+1)}_i[E_l^{(I+1)}] x^iR_{\beta}(x), 
\end{equation}
define good, approximate, solutions to the physical wave functions. 
The same should hold for the
corresponding, renormalized, truncated power series expansion
\begin{equation}
\Psi_l(x) \approx \sum_{i=0}^{I+1} {a_i[E_l^{(I+1)}]\over 
{a_{I+1}[E_l^{(I+1)}]}} x^iR_{\beta}(x).
\end{equation}
Note that each of the expansions is 
normalized by setting the highest degree coefficient to unity.

For fixed, and sufficiently large, values of $I+1$, 
the preceding partial sums are expected to agree, 
term by term, for each of the roots $E_{l}^{(I+1)}$.
In particular, based on this concurrence, 
one expects ${a_I[E]\over {a_{I+1}[E]}} \approx V^{(I+1)}_I [E]$,
which upon comparing with  Eq.(10) leads to the desired conclusion:
\begin{equation}
{a_I[E_l^{(I)}]\over {a_{I+1}[E_l^{(I)}]}} 
\approx V^{(I+1)}_{I}[E_{l}^{(I)}] = \infty,
\end{equation}
or 
\begin{equation}
a_{I+1}[E_l^{(I)}] = 0,
\end{equation}
and $\Psi_l(x) \approx {\cal N}\sum_{i=0}^{I}a_i[E_l^{(I)}]x^{i}R_{\beta}(x)$.

We now demonstrate the capabilities of the preceding method.
For the case of exactly solvable models, 
e.g., $V(x)= x^2$ or $V(x)= x^2 + g/x^2$,
our approach readily yields 
the exact solutions, once the proper reference 
functions are selected: $R_\beta = \exp (-x^2/2)$, 
and $R_\beta = x^{\alpha} \exp (-x^2/2)$ ($\alpha=(1+\sqrt{1+4g})/2$), 
respectively. 

Let us now consider the quartic anharmonic oscillator,
$V(x)= x^2 + g x^4$. Using $R_\beta = \exp (-\beta x^2)$ one obtains the
recursion relation:
\begin{equation}
a_n(E) = { \Omega_n\ a_{n-2}(E)+(1-4\beta^2)a_{n-4}(E) 
+ g a_{n-6}(E)  \over n(n-1)}\ ,
\end{equation}
where $\Omega_n = 4 \beta n-6\beta-E$ and $a_n=0$ for $n<0$.  
Table I shows the calculated energies of 
the ground and first excited states. Our method shows systematic
convergence for increasing $I$, exceeding some of the high accuracy 
solutions published [6 - 8].
Figure 1 shows the dependence of the ground state energy on the
parameter $g$. 
Figure 2 shows the corresponding wave functions.

An important version of the quartic anharmonic oscillator potential is
the double well problem $V(x) = -Z^2x^2 + x^4$. 
It is well-known that in the deep well limit ($Z^2 \rightarrow \infty$), 
the two lowest states are almost degenerate. Application of 
our method (refer to Table II)
readily confirms this, and by its high accuracy nature, 
significantly disagrees with the predictions of
de Saavedra and Buendia (SB) [9]. In particular, 
for $Z^2=25$, we observe that
the quasi-degenerate nature of the 
ground and first excited state energies
becomes apparent only after 26 significant digits, 
not the 16 predicted by SB.

Similar calculations have been carried out 
for higher degree potentials such as the
sextic, octic, and dectic
anharmonic potentials. The results are given in Table III.

The generality of our method permits the  study of large
classes of problems. In particular,
transcendental potentials can be analyzed, 
provided the potential function, $V(x)$,
admits a power series expansion which is 
monotonically convergent (non-alternating).
For instance,  in the case of
$V(x) = \exp(x^2)-1 $,  
we can readily obtain the first three energy levels:
$E_0 = 1.356371240$, 
$E_1 = 4.633078503$, and 
$E_2 = 8.9706782$. 
($R_{\beta}(x) = e^{-x^2}$, and $a_n$ generated up to $n \leq 80$).

Another type of potential which can be investigated is
$V(x) = V_0(x_0 + x)^{\alpha}$, 
where $\alpha > -2$. For the two cases 
$-2 <\ \alpha < 0$ and $0 < \alpha < \infty$
the asymptotic forms for the wave function can be 
studied via 
$\Psi(x) = 
\phi_{-}(x) \ e^{-\sqrt{-E} x}$ and $\Psi(x) = \phi_{+}(x) \ e^{-\beta x^2}$,
respectively. Upon making the transformation $y = \sqrt{
\log(1+x /x_0)}$, one 
can readily obtain the corresponding 
differential equations for $\phi_{\pm}(y)$, and proceed
to generate the necessary power series expansion, 
$ \phi_{\pm}(y) =\sum a^{(\pm)}_n y^n$.
In Figure 3 we plot $E[\alpha]$ vs. $\alpha$ for the 
cases of \{$x_0=1$, $V_0=-2$; $-2< \alpha < 0$\} 
and \{$x_0=1$, $V_0=1$; $0<\alpha < 4$\}.
%
% Momentum Space
%

The selection of the reference function is clearly important.
In the case of the sextic anharmonic oscillator ($V(x) = x^2 + gx^6$), 
our method works for the 
reference functions
$e^{-\beta x^{\sigma}}$, $\sigma = 2, \ {\rm and}\ 3$.
For the case $\sigma = 4$ (and $\beta = 
{\sqrt{g}\over 4}$), which corresponds to 
the asymptotic form 
of the wave function, 
no convergent roots were observed upon solving Eq. (14). 

Quantization is a global problem, 
not a local one.  Configuration space analysis, by its manifestly local
nature, may not always be a suitable 
representation in which to quantize. However, a momentum space
representation is more appropriate. 
Specifically, the power series expansion for the Fourier transform
of the wave function, 
${\hat \Psi}(k) = \int dk\ e^{-ikx} \Psi(x) = 
\sum_{p=0} {{(-ik)^p\over {p!}}} \mu(p)$, involves
global quantities, the (Hamburger) 
power moments, $\mu(p) = \int dx \ x^p \Psi(x)$. As such, one
would expect that our method would be 
more effective in momentum space. These observations are at the heart
of various moment based quantization approaches [2], 
particularly the Eigenvalue Moment Method (EMM) [10].

Extending our method into momentum
space presents additional features not 
encountered in configuration space. The most important of these  is
that more variables (the {\it missing moments} 
within the EMM approach)
are encountered, regardless of the 
spatial dimension, $\cal D$, of the problem. Solving spatial problems
of dimension ${\cal D} = 1$ 
presents similar theoretical/algorithmic 
challenges to those in higher dimensions. We outline
the essentials.

Limiting ourselves to symmetric $\Psi(x)$ configurations, 
for simplicity, the
coefficients of the power series 
expansion for the momentum space wave function,
$\hat \Psi(k) = \int dk\ e^{-ikx} 
\Psi(x) = \sum_{\rho=0}{{(-k^2)^{\rho}}\over {(2\rho)!}}u(\rho)$ 
satisfy a linear, moment recursion equation, 
resulting in the linear relations 
$u(\rho) = \sum_{\ell=0}^{m_s} M_E(\rho,\ell) 
u(\ell)$, $0 \leq \rho <\ \infty$; the $ M_E(\rho,\ell)$'s 
are known, and the {\it missing moment}
order, $m_s$, is problem dependent [10]. 

Implementing our quantization procedure on the representation
$\hat \Psi(k) = (\sum_{n=0}a_{2n}k^{2n}) e^{-\beta k^2}$, we obtain
\begin{equation}
a_{2n}[E,u(0),...,u(m_s)] = 
\sum_{\ell=0}^{m_s}D_{n,\ell}[E] u(\ell),
\end{equation}
where 
\begin{equation} D_{n,\ell}[E] = \sum_{\rho_1 + \rho_2 = n} 
{{(-k^2)^{\rho_1}M_E(\rho_1,\ell) 
\beta^{\rho_2}}\over
{(2\rho_1)! \rho_2 !}}.
\end{equation}
In principle, there will be a sequence 
of energy and missing moment values satisfying
$a_{2n}[E^{(n)},\{u^{(n)}(\ell)\}] = 0$ 
converging  to the respective physical state values as $n \rightarrow \infty$. 
Since the $D_{n,\ell}[E]$'s are not
expected to define a degenerate matrix for all $E$'s, 
as $n \rightarrow \infty$, we can 
approximate the converging sequence by
considering  the $[m_s+1] \times [m_s+1]$ matrix equation
\begin{equation}
\sum_{\ell_2 = 0}^{m_s} D_{n+\ell_1,\ell_2}[E] u(\ell_2) = 0,
\end{equation}
and the ensuing determinant equation,
\begin{equation}
Det\Big(D^{(n)}[E] \Big) = 0.
\end{equation} 

Implementation of 
this for the quartic ($m_s = 1$) and sextic ($m_s = 2$)
anharmonic oscillators yielded results
consistent with those cited in Tables I and III. 

Some problems can involve no missing moments.
One of these is the aforementioned 
sextic anharmonic oscillator ($m_s = 0$ formulation), 
provided one implements
the above formalism with respects to the 
configuration space expression $\tilde \Psi(x) = 
\Psi(x) e^{-{\sqrt{g}\over 4}x^4}$. 

The same holds for 
the problem $V(x)= x^2 + {{gx^2}\over
{1+\lambda x^2}}$, provided 
 $\tilde \Psi(x) = {\Psi(x)\over {1+gx^2}}e^{-{1\over 2}x^2}$ [11]. 
Table IV summarizes our results for this case, which
surpass
the accuracy 
calculated through an analytic continuation quantization
procedure [12].

In summary, we have developed a nonperturbative approach for the inner
product quantization procedure which allows one to calculate the
energies and wave functions of the Schr\"odinger equation. 
We have demonstrated that our method yields 
excellent numerical results for various
quantum mechanical problems. A full account of the 
approach, including the application to higher
dimensional models and resonant states, will be published elsewhere.

This work was supported in part by the National Science Foundation
under Grant No. HRD9450386,
Air Force Office of
Scientific Research under Grant No. F49620-96-1-0211, and Army
Research Office under Grant No. DAAH04-95-1-0651.

%
% 	References
%

%
%	Tables and Figures
%
\begin{table}
\caption{The calculated ground and 
first excited state energies for the quartic 
anharmonic oscillator with $g=1$.}
\begin{center}
\begin{tabular}{lccl}
\multicolumn{1}{l}{$I$} &
\multicolumn{1}{c}{$\beta$} &
\multicolumn{1}{c}{$n$} &
\multicolumn{1}{l}{$E_n$} \\ \hline
10  & $1/2$ & 0 & 1.41 \\
      &                & 1 & 4.9   \\
      &              1 & 0 & 1.392 \\
      &                 & 1 & 4.65 \\
40  & $1/2$ & 0 & 1.392 349 \\
      &                 & 1 & 4.648 84  \\
      &              1 & 0 & 1.392 351 641 4\\
      &                 & 1 & 4.648 812 70\\
160 & $1/2$ & 0 &  1.392 351 641 530 291 \\
      &                 & 1 &  4.648 812 704 212 \\
      &              1 & 0 & 1.392 351 641 530 291 855 657 507 876 \\
      &                 & 1 & 4.648 812 704 212 077 536 377 032 91 \\ 
\multicolumn{2}{l}{Refs. [4-6]} & $E_0$ & 1.392 351 641 530 291 85 \\
\multicolumn{2}{l}{} & $E_1$ & 4.648 812 704  \\
%       
%10  &  0.5  & 1.41                           \\
%40  &  0.5  & 1.392 349                     \\
%160 &  0.5 & 1.392 351 641 530 291  \\
%\hline
%10  &  1.0 & 1.392  \\
%40   &  1.0 & 1.392 351 641 4    \\
%160  &  1.0 &  1.392 351 641 530 291 855 657 507 876 \\
%\hline
%\multicolumn{1}{l}{$I$} &
%\multicolumn{1}{c}{$\beta$} &
%\multicolumn{1}{l}{$E_1$} \\ \hline
%10   &  0.5 & 4.9                         \\
%40   & 0.5 & 4.648 843               \\
%160 & 0.5 & 4.648 812 704 212   \\
%\hline
%10    &  1.0 & 4.65    \\    
%40    & 1.0 &  4.648 812 70 \\
%160  & 1.0 &  4.648 812 704 212 077 536 377 032 91\\
%\hline
\end{tabular}
\end{center}
\caption{The calculated ground and first excited state energies 
for the potential $V(x) =-Z^2 x^2+x^4$.}
\begin{center}
\begin{tabular}{lcr}
\multicolumn{1}{l}{$Z^2$} &
\multicolumn{1}{l}{$Parity$} &
\multicolumn{1}{c}{$E_{\pm}$} \\ \hline
0    & + &     1.060 362 090 484 182 899 647 046 016  \\
      & $-$ &     3.799 673 029 801 394 168 783 094 188 \\ 
1    &  +&     0.657 653 005 180 715 123 059 021 723 \\
      &  $-$ &     2.834 536 202 119 304 214 654 676 208 \\
5    &  +&  -3.410 142 761 239 829 475 297 709 653 \\
      &  $-$&  -3.250 675 362 289 235 980 228 513 775 \\
10   &  +& -20.633 576 702 947 799 149 958 554 634\\
       &  $-$& -20.633 546 884 404 911 079 343 874 899\\
15   &  +&  -50.841 387 284 381 954 366 250 996 515 \\
      &  $-$&  -50.841 387 284 187 005 154 710 149 735 \\ 
25  &  +& -149.219 456 142 190 888 029 163 966 538 \\ 
      &  $-$& -149.219 456 142 190 888 029 163 958 974 \\
\end{tabular}
\end{center}
\caption{The calculated ground state energies of the sextic, octet and dectic 
anharmonic potentials for $g=1$.}
\begin{center}
\begin{tabular}{lll}
\multicolumn{1}{c}{$V(x)$} &
\multicolumn{1}{c}{$E_0$ (Ref. [4])} &
\multicolumn{1}{c}{$E_0$} \\ \hline
$x^2+x^6$     & 1.435 624 619 0 & 1.435 624 619 003 392 231 569 \\
$x^2+x^8$     & 1.491 019 895    &  1.491 019 895 662 \\
$x^2+x^{10}$ &                   & 1.546 263 512 6 \\
\end{tabular}
\end{center}
\caption{The first four symmetric state energies for 
the rational fraction potential 
$V(x)= x^2 + {{gx^2}\over {1+\lambda x^2}}$ for $\lambda=g=0.1$}
\begin{center}
\begin{tabular}{cl}
\multicolumn{1}{l}{$n$} &
\multicolumn{1}{c}{$E_n$} \\ \hline
0  & 1.043 173 713 044 445 233 778 700 870 546 094 \\
2  & 5.181 094 785 884 700 927 110 409 072 888 3 \\
4  & 9.272 816 970 035 252 254 582 438 478 9 \\
6  & 13.339 390 726 973 551 232 933 170 5 \\
\end{tabular}
\end{center}
\end{table}

\noindent
\figure{FIG 1. The calculated ground state energy for the 
quartic anharmonic oscillator.}
\noindent
\figure{FIG 2. The calculated ground and first excited state 
wave functions for the quartic anharmonic oscillator. }
\noindent
\figure{FIG 3. The calculated ground and first excited state energies for 
the potential $V(x) = V_0 (x+x_0)^{\alpha}$ for $-2<\alpha<2$. }
\end{document}